# Extraordinary strain hardening from dislocation loops in defect-free Al nanocubes

Mehrdad T. Kiani[a], Zachary H. Aitken[b], Abhinav Parakh[a], Yong-Wei Zhang[b], X. Wendy Gu[c,1]

[a]Department of Materials Science & Engineering, Stanford University, Stanford, CA 94305; [b]Institute of High Performance Computing, A*STAR, Singapore 138632; [c]Department of Mechanical Engineering, Stanford University, Stanford, CA 94305

**The interaction of crystalline defects leads to strain hardening in bulk metals. Metals with high stacking fault energy (SFE), such as aluminum, tend to have low strain hardening rates due to an inability to form stacking faults and deformation twins. Here, we use *in situ* SEM mechanical compressions to find that colloidally synthesized defect-free 114 nm Al nanocubes combine a high linear strain hardening rate of 4.1 GPa with a high strength of 1.1 GPa. These nanocubes have a 3 nm self-passivating oxide layer that has a large influence on mechanical behavior and the accumulation of dislocation structures. Post-compression TEM imaging reveals stable prismatic dislocation loops and the absence of stacking faults. MD simulations relate the formation of dislocation loops and strain hardening to the surface oxide. These results indicate that slight modifications to surface and interfacial properties can induce enormous changes to mechanical properties in high SFE metals.**

*in situ* scanning electron microscopy | yield strength | plasticity

Strain hardening in metals is intimately related to the accumulation and interaction between dislocations and other crystalline defects(1). The stacking fault energy (SFE) is the energetic cost of disrupting the stacking sequence in close-packed crystals, which is an intrinsic material parameter commonly used to predict plastic behavior and strain hardening in fcc metals. Low SFE metals can easily form partial dislocations which lead to stacking faults and deformation twins. These defects enhance strain hardening by interacting with mobile dislocations. Aluminum is a low density and high SFE metal in which it is energetically unfavorable to form stacking faults or twins. Partial dislocations in Al are limited to ~20 nm in length(2), which prevents interactions with other defects.

Recently, unusual deformation microstructures have been observed in extruded single crystal Al and Al nanowires which conflict with conventional knowledge of high SFE metals. In bulk single crystal Al samples that undergo equal channel angular pressing, it is found that a crystal orientation that maximizes shear stress for a single twinning system can induce the formation of short, ~10 nm stacking faults and microtwins with thickness of several atomic layers(3). <110>-oriented Al nanowires tested in tension formed twins in a zig-zag pattern along the length of the nanowire. This led to a high strength of 2.7 GPa but limited plasticity, because twinning resulted in strain localization(4, 5). Focused ion beam (FIB) milled <100>-Al nanowires initially deformed via the formation of prismatic dislocation loops(6). Further deformation occurred via motion of pre-existing threading dislocations. A high final dislocation density of $10^{13}$-$10^{14}$ m$^{-2}$ indicates that most dislocations did not annihilate at free surfaces despite strong image forces.

While these previous studies clearly showed the strong influence of initial defect structure, crystal orientation, and surface properties on the deformation in low SFE metals, these unusual deformation mechanisms, and the challenge of achieving both high strength and strain hardening in metallic nanostructures(7, 8), motivate further investigation into nanoscale Al. Here, we use *in situ* scanning electron microscope (SEM) to compress defect-free <100>-oriented Al

> **Significance**
>
> **Low stacking fault energy (SFE) metals tend to have higher increased strain hardening through stacking fault and twin formation. This relationship between SFE and strain hardening may break down in nanostructured and nanoscale metals. In this work, we compress defect-free Al nanocubes, a high SFE metal, which display pronounced strain hardening not seen in other defect-free nanocubes. Post-compression TEM shows stable dislocation loops with no stacking faults or twins. From MD simulations, we relate strain hardening to the thin self-passivating oxide, which prevents dislocation annihilation at free surfaces. These findings provide new insight into the role interface engineering could play in enhancing mechanical properties of high SFE metals.**

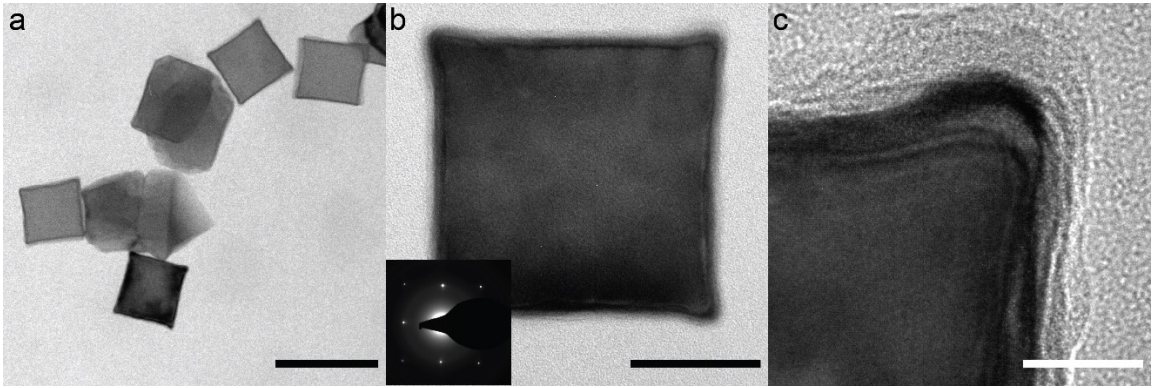

**Figure 1.** a) Representative Al nanocubes. Scale bar is 200 nm. b) Single nanocube with inset showing corresponding TEM diffraction. Scale bar is 50 nm. c) HRTEM of nanocube corner in b) showing oxide protective layer. Scale bar is 5 nm.

nanocubes made using colloidal synthesis. The nanocubes are coated with a nanoscale oxide layer which results in well controlled surface properties. Stress-strain curves of individual nanocubes of 114 nm in size showed pronounced strain hardening of 4 GPa, high yield strength of 1.1 GPa and suppressed dislocation avalanches compared to other fcc nano and microstructures. Post-compression TEM imaging show the formation of stable prismatic dislocation loops with no stacking faults or twins. Molecular dynamics (MD) simulations reveal the role of surface roughness and the rigid surface oxide on the deformation mechanism, mechanical behavior and strain hardening. The combination of high strength and strain hardening in these nanomaterials through surface and interfacial modification provide a new route to design nanostructured materials.

## Results

Synthesized Al nanocubes are shown in Figure 1a. The nanocube yield in the synthesis was approximately 40% with a side length of 114.3±5.5 nm. All Al nanocubes were single crystalline. No dislocations, twins, or other defects were seen in TEM images of nanocubes (Figure 1b). Upon exposure to air, the Al nanocubes form a self-passivating amorphous $Al_2O_3$ layer of 2-4 nm(9) (Figure 1b).

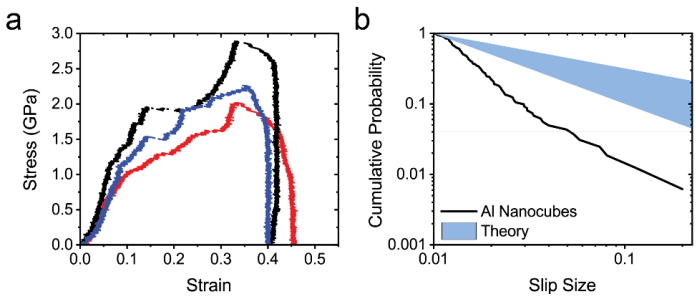

**Figure 2.** a) Representative compressive engineering stress-strain curves. b) Cumulative probability plot of slip sizes compared to theoretical predictions for single crystal fcc nanostructures(12).

15 Al nanocubes were compressed using an *in situ* mechanical tester in a SEM. Representative engineering stress-strain curves are shown in Figure 2a. The yield strength was 1120± 300 MPa, which is approximately double what is expected from FIB-milled Al nanopillars of a similar size(10). Yielding occurs through dislocation nucleation in defect-free metallic nanostructures(11), which is the case here. In contrast, FIB-milled micro/nanopillars have pre-existing dislocations, and yield when these dislocations start to move, which requires a lower stress. Upon yielding, small intermittent strain bursts are observed in the nanocubes which are indicative of the dislocation avalanches typical of metallic nanostructures(12).

Pronounced strain hardening occurs, which was quantified using a linear strain hardening rate, $\Theta = \frac{d\sigma}{d\epsilon}$. The Al nanocubes exhibited a hardening rate of 4.1 ± 1.6 GPa. The true cross-sectional area cannot be measured *in situ*, so the true strain hardening rate is estimated using the Poisson's ratio of Al(13), and found to be 2.1±1.1 GPa. In comparison, bulk <100> single crystal Al has a negligible hardening rate (<50 MPa)(14). Other defect-free Ag and Cu nanocubes and single crystal fcc and bcc nanopillars show no strain hardening on compressive engineering stress-strain curves(12, 15, 16). To further quantify the slip behavior, the cumulative probability of slip sizes for all compression tests was plotted in Figure 2b. This curve has a slope of approximately -2. Theoretical and experimental studies of fcc nanopillars have a slope of -0.5 to -1 while experimental studies of Ag and Cu nanocubes have slopes of -1.6 and -1 respectively(12, 15). The large negative slope for Al nanocubes indicates that larger dislocation avalanches are being suppressed during compression.

Post-compression TEM images of a representative Al nanocube are shown in Figure 3. This was obtained by dropcasting nanocubes on a FIB-milled Si liftout grid, compressing the nanocubes *in situ* SEM, and then transferring the liftout grid to the TEM for imaging. Figure 3a shows the flattened Al nanocubes which are slightly tilted away from the <110> orientation with respect to the electron beam. Slip offsets are not observed on the nanocube surface, which would be indicative of dislocation avalanches that nucleate from the free surface. Instead, we observed small dark spots scattered through the Al nanocube. A <100>-oriented Al nanocube has a similar appearance (Figure S1). At a higher magnification, lines of contrast are observed along {111} planes inside the <110>-oriented nanocube (Figure 3c). The dark spots seen in Figure 3a are associated with regions near the lines of contrast. These lines of contrast, which range in length from 5-15 nm, begin and end inside the nanocube and do not terminate at a free surface. HRTEM imaging shows that the lines of contrast are not stacking faults (Figure 3c). Fast Fourier transforms (FFTs) of the regions of contrast show variance in the {111} plane spacing, which does not appear in the defect-free region (Figure 3d).

From the combination of TEM images, FFT analysis, and previous experimental results on (100)-oriented FIB milled Al nanostructures(6), these lines of contrast are identified as prismatic {111} dislocation loops. We are unable to perform g·b analysis due to limitations in tilt angle due to the wedge-shaped substrate and the high density of defects. However, we determine that $b = {^a/_2} <110>$ due to the absence of an extra half plane or stacking fault, which would be indicative of a Frank partial loop. The dislocation density is estimated to be $\sim 10^{15}$ m$^{-2}$ by calculating the line density from Figure 3b and assuming circular dislocation loops. This is one to two orders of magnitude higher than previously observed in single crystal Al nanostructures under compressive or tensile testing(6, 17).

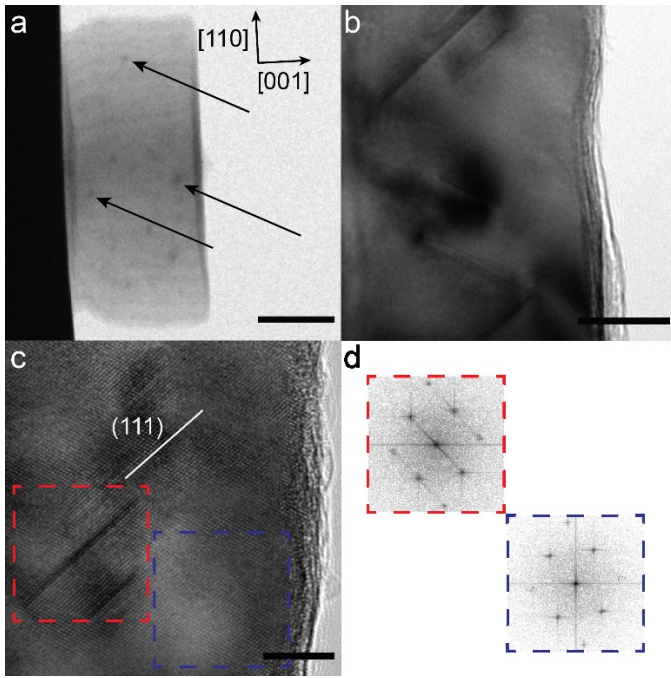

**Figure 3.** a) Compressed Al nanocube which is tilted slightly off <110> orientation. Arrows point to spots of contrast. Scale bar is 50 nm. b) Magnified region showing lines of contrast along the {111} planes. Scale bar is 10 nm. c) HRTEM image of dislocation loop along the (111) plane. Scale bar is 5 nm. The FFTs of the dotted red and blue region are shown in d).

The statistical likelihood for dislocation interactions decreases in nanoscale samples due to decreased sample volume and increased dislocation annihilation at free surfaces. This has led to minimal strain hardening and large slip events in defect-free fcc Ag(15, 18), Cu(15), Au@Ag(19), and Ni$_3$Al(20) nanocubes. In contrast, the large strain hardening rate and presence of dislocation loops in compressed Al nanocubes indicate that dislocation interactions occur in the small sample volume. Strain hardening mechanisms such as stacking fault-dislocation interactions(21) or crystal rotation(22) can be ruled out given the absence of stacking faults in TEM images (Figure 3c) and the <100> loading orientation of Al nanocubes, respectively.

It is unclear whether the dislocation loops are the primary driver of plasticity or are a byproduct of other dislocation interactions. Prismatic dislocation loops have been observed to nucleate in single crystal <100>-oriented Al nanostructures in tension but were transient and disappeared with increased strain(6). The proposed mechanism was dislocations loop formation via non-local homogenous nucleation(23, 24). For a 10 nm dislocation loop, the uniaxial stress required for non-local loop nucleation is approximately 3.5 GPa, which cannot be achieved under most conditions(24). Dislocation loops were also formed during indentation of a sharp tip into a surface(25–27). The stress concentration underneath the tip leads to the nucleation of half loops which cross slip to form stable prismatic dislocations. An analogous process could play a role in our study, in which the nanoscale surface roughness of both the flat punch tip and the Al nanocube surface lead to stress concentrations that induce loop formation.

Al nanocubes differ from other colloidally synthesized fcc nanocubes in that they form a native oxide which serves as a rigid surface that can block dislocation motion at the interface. However, previous work has shown conflicting evidence on the role of a coating. Compression of FIB-milled pillars with sputtered or atomic layer deposited metal or oxide coatings showed strain hardening with final dislocation densities comparable to the Al nanocubes due to the inability of dislocations to annihilate at free surfaces(28, 29). In contrast, Au nanowires coated with alumina had no change in plastic behavior(30). Previous MD simulations of 4-7 nm Al nanowires with Al$_2$O$_3$ coating show slight hardening only in compression due to increase in Al-O bonds(31, 32). These MD simulations had an oxide volume fraction of ~50%, which is significantly different from the oxide volume fraction of ~8% in the current study.

MD simulations of 40 nm Al nanocubes were performed to understand the role of surface roughness and the oxide layer on dislocation loop formation and strain hardening. Figure 4 shows the stress-strain curve and defect structure of a rough nanocube, modeled with 2 nm radius hemispherical protrusions on the surface. This nanocube does not have a rigid surface. The size of these features is comparable to the surface roughness seen in TEM images (Figure 1c) and can lead to stress concentrations. As compared to simulations of a smooth nanocube (Figure S2), the elastic slope of the rough nanocube (12 GPa) and the peak stress (0.96 GPa) are significantly reduced. The reduction in elastic modulus is indicative of a non-elastic response of the rough surface features; the surface hemispheres are compressed before the entire cube responds elastically. Inspection of the spherical protrusions indicates activation of partial dislocation nucleation at strains below 8.8% which is the strain at peak stress. Following the peak stress, plastic deformation occurs with a hardening rate of 1.3 GPa between strains of 10% to 18%. Figure 4f shows the partial dislocation structures that nucleate from the spherical protrusions. Due to interactions between different slip systems, these dislocations remain confined near the surface. Upon further straining, dislocation loops nucleate from these partial structures as shown in Figures 4g and 4h. An example of the resulting loops that nucleate into the bulk of the nanocube is shown in Figure 4i.

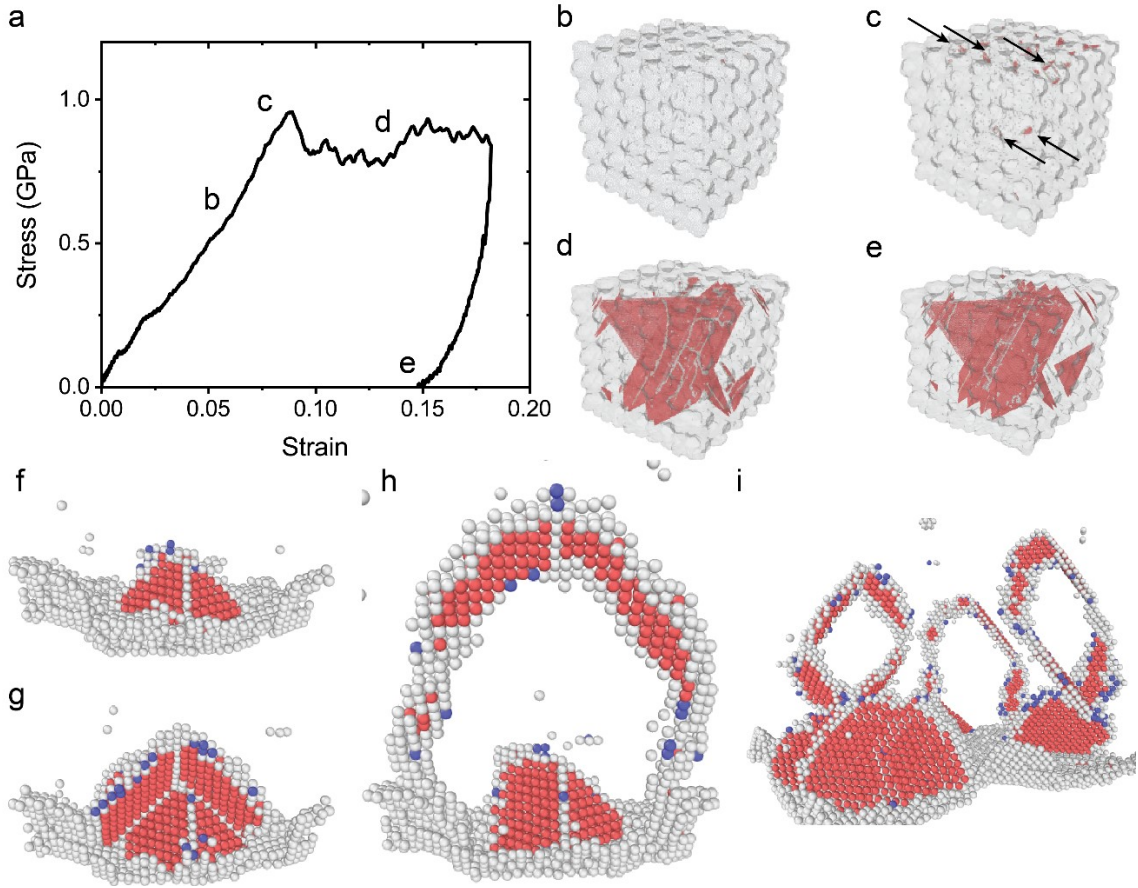

**Figure 4.** Simulation results from compression of rough nanocube without a rigid surface. a) Stress-strain data. Atomistic snapshots of deformed structure at strains of b) 0.06, c) 0.08, d) 0.13, and e) after unloading. f)-h) Formation of dislocation loop that occurs during the shaded region of the stress-strain curve. Arrows in c) point to nucleation sites for dislocation loops along the top and bottom surfaces. i) Nucleation of several dislocation loops at 8.8% strain. Only non-FCC atoms are shown (i.e. defected or surface atoms), with specific CNA classification of red (HCP), blue (BCC), and white (unknown).

Figure 5 shows the stress-strain curve and defect structure of a smooth nanocube with a rigid surface which mimics the surface oxide layer in the experimental nanocubes. The elastic slope is 65 GPa and the maximum stress is 4.25 GPa. This indicates that the additional stiffness of the rigid surface boundary conditions leads to a significant strengthening effect. Plastic deformation occurs with a hardening rate of 3.9 GPa from 10% to 16% strain, which is significantly greater than the non-rigid samples and approaches the hardening rate of the experimental nanocubes. Figure 5f shows how dislocations pile-up at the surface, unable to penetrate due to the imposed rigid boundary conditions. Deformation is dominated by emission of partial dislocations with extended stacking faults.

## Discussion

MD simulations suggest that the mechanical behavior of the Al nanocubes is due to the combined effect of surface roughness and the rigid oxide layer. The introduction of hemispherical surface roughness results in the appearance of dislocation loops similar to those observed experimentally. These defect structures are absent in the MD samples with perfectly smooth surfaces (Figure S2). The rigid surface boundary condition that mimics the presence of a native oxide results in a significant hardening effect. The strain hardening rate of the Al nanocube with the rigid surface was 3.9 GPa, a more-than-two-fold increase as compared to samples without the rigid surface. Inspection of the defect structure in this sample shows dislocations piled-up at the rigid surface. These MD simulations suggest possible origins to the dislocation loops and high hardening rate observed experimentally. Considering these results and previously reported

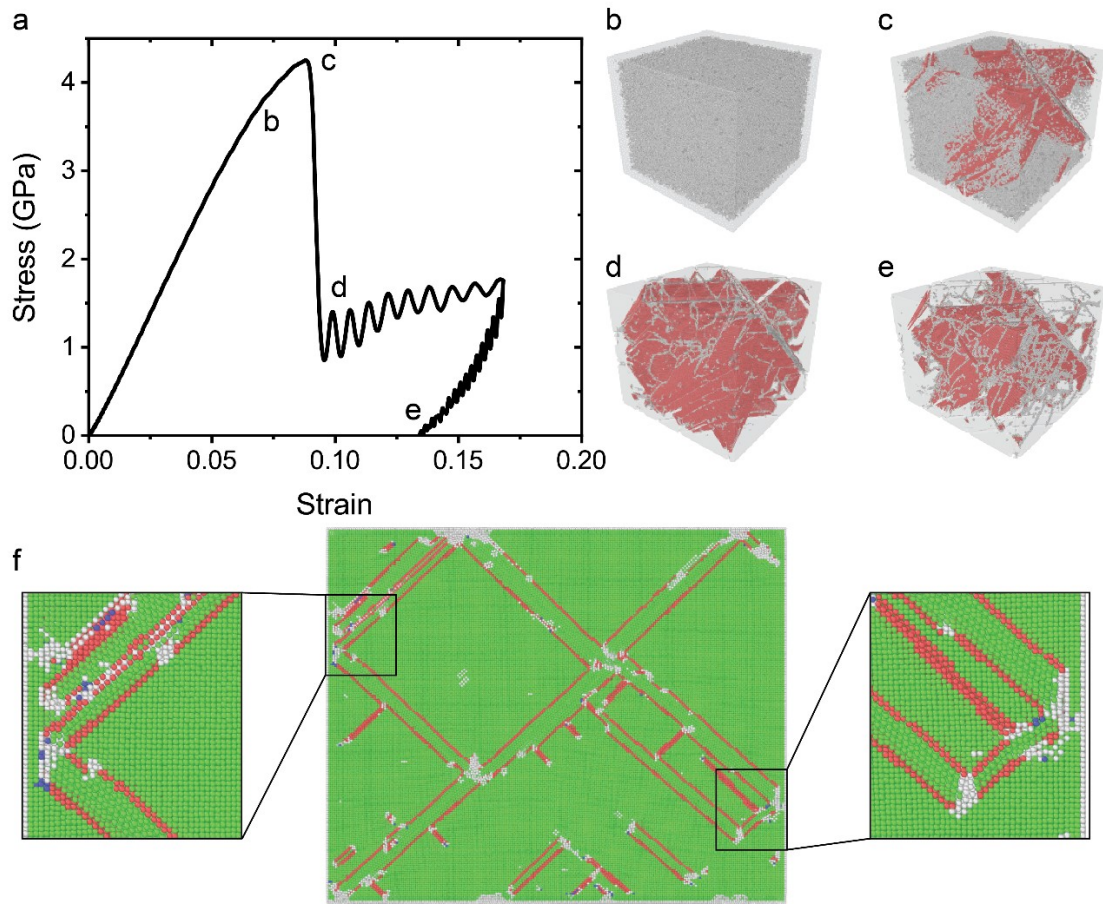

**Figure 5.** Simulation results from compression of smooth nanocube with a rigid surface. a) Stress-strain data. Atomistic snapshots showing the deformation structure at strains of b) 0.086, c) 0.09, d) 0.096, and e) 0.13. f) Dislocation pile-ups at the rigid surface that are unable to exit the nanocube. Specific CNA classification are green (FCC), red (HCP), blue (BCC), and white (unknown).

simulations(26), it appears that surface stress concentrators can lead to partial dislocation interactions near the surface and subsequent formation of loops in the bulk of the nanocube. The post-deformation presence of loops in the Al nanocube may be caused by the trapping of dislocations by the rigid surface oxide. This leads to increased hardening as dislocations accumulate, which promotes dislocation interactions.

Stacking faults are formed in the MD simulations but not the experiments because of differences in nanocube size and strain rate. Dislocation activation volume scales with $b^3$, where $b$ is the magnitude of the Burgers vector. This indicates that partial dislocations are preferable over full dislocations because partial dislocations have a lower Burgers vector of $b = a/6 \langle 211 \rangle$(11). This preference for partial dislocations is exaggerated in small volumes since the leading partial will annihilate on the other side before a trailing partial, leaving a stacking fault. Previous MD simulations of nanocrystalline Al show a grain size dependent transition in stacking fault formation(33). Below ~45 nm grain size, the high strain rate coupled with the small sample volume leads to stacking fault formation in which the leading partial annihilates at the opposite side of the grain before nucleation of the trailing partial. Above this grain size, the trailing partial nucleates before the leading partial annihilates, thus preventing stacking fault formation. This size dependent transition in deformation mechanism is similar to the nanocubes, which can be thought of as a single grain within a rigid grain boundary formed by the oxide layer. Small nanocubes at high strain rates (MD simulation) show the formation of stacking faults, while the larger experimental nanocubes do not. In addition, when comparing the results from the smooth nanocube without (Figure S2) and with a rigid surface (Figure 5), we observe that dislocation interactions at the surface drive strain hardening and not stacking fault-dislocation interactions.

In summary, *in situ* SEM compressions of 114 nm defect-free Al nanocubes showed extensive strain hardening that is two order of magnitudes higher than bulk single crystal Al. Numerous dislocation loops were observed in post-compression TEM images, as well as the absence of stacking faults and twins. MD simulations of 40 nm Al nanocubes show that dislocation loops form due to stress concentrators on the oxide coating while strain hardening occurred due to trapping of dislocations by the oxide coating. These results provide strong evidence that high SFE metals can have high strain hardening through dislocation accumulation. This can be achieved through careful control of surfaces and interfaces, such as the surface oxide in the nanocubes, or grain boundaries in polycrystalline metals.

## Materials and Methods

**Synthesis.** All chemicals were purchased from Sigma-Aldrich. 0.5 M dimethylethylamine alane (DMEAA) in toluene, 0.5 M Tebbe's reagent in toluene, cyclohexane, and tetrahydrofuran (THF) were stored under in a glovebox under argon. All dilutions from as-purchased concentrations were performed in a glovebox using a glass syringe.

Al nanocubes were synthesized using a previously developed method(9). 200 mM $AlH_3$ solution was made by adding 4 mL of DMEAA to 4 mL of THF in a glovebox. After transferring to a Schlenk line under argon, the solution was heated to 70°C and 2 mL of 50 mM Tebbe's reagent was injected with a glass syringe. The solution was stirred for approximately 2 hours before quenching with cyclohexane. The solution was centrifuged at 5000x*g* for 5 minutes and then redispersed in IPA. Centrifugation was repeated twice more at 10000x*g*.

**Mechanical Testing & Characterization.** Al nanocubes were dropcast onto a (100) Si wafer before testing. Compression tests were performed using the Nanoflip *in situ* mechanical tester (Nanomechanics Inc.) inside a FEI Helios NanoLab 600i DualBeam FIB/SEM. Compression tests were performed under load control at .001 mN/s using a diamond flat punch tip. To generate cumulative probability curves, all slip events larger than 0.01 strain units were individually measured from generated stress strain curves. To calculate the slope of the cumulative probability plots, $\alpha$, we use Equation 1(34), where $n$ is the number of slip events, $s$ is the slip magnitude, and $s_{min}$ is the smallest slip event measured:

$$\alpha = n \left( \sum_{i=1}^{n} \ln \frac{s_i}{s_{min}} \right)^{-1}$$

TEM images were acquired on a FEI Tecnai G2 F20 X-TWIN TEM at 200 kV. FFT images were acquired using ImageJ.

**Simulations.** Molecular dynamics simulations were performed using LAMMPS(35). A 40 nm Al cube was initialized between two rigid plates with [100] directions aligned with the orthogonal simulation box directions. Prior to compression, the structure was relaxed for 100 ps under an NVT ensemble followed by 100 ps of Langevin dynamics to dampen pressure oscillations. Following experimental methodology, a force-controlled loading approach was used for simulation. A loading rate of $10^5$ μN/s was applied to the top plate which results in a variable displacement rate with a maximum of $9*10^8$ s$^{-1}$ by the end of compression. We note that although simulations were run by applying a linear force to the compression plate, without a feedback loop to adjust the applied force there may still be stress drops in the stress-strain signature. A modified embedded-atom method (MEAM) potential was used which shows excellent agreement with the equilibrium lattice structure, surface energies, elastic properties and generalized stacking fault energy curve of aluminum(36). Visualization of the structure and dislocation analysis was done via common neighbor analysis (CNA) and the dislocation extraction algorithm (DXA) as implemented in the OVITO visualization software(37).

Two types of nanocube samples along with a reference nanocube were used to investigate the effect of sample geometry and boundary conditions. The reference nanocube was a perfectly smooth cubic geometry with unconstrained degrees of freedom at the surface. To investigate the effects of surface roughness, a cubic sample with semi-spherical protrusions and indentations was constructed. The diameter of the protrusions/indentations was 40 Å and they were placed in a regular grid with alternating protrusions and indentations to cover the entirety of the cube's surface. To investigate the effect of a rigid oxide surface layer, a nanocube with rigid surface boundary conditions was also subjected to compression simulations. This sample was initially smooth and the first 4 atomic layers from the free surface were considered the surface layer. Surface atoms of a given face were required to move as a rigid body in the direction perpendicular to that face. This disallowed shear across the face, mimicking a rigid surface layer. Atoms belonging to 2 surface faces (edge atoms) or belonging to 3 surface faces (corner atoms) were subject to the rigid body condition of all adjacent faces. This means that edge atoms must move as a rigid body in the directions perpendicular to its axis and corner atoms must move as a fully rigid body. This set of boundary conditions allows the box to undergo arbitrary deformation that maintains its rectangular shape while disallowing shear across any face.

**ACKNOWLEDGMENTS.** We gratefully acknowledge financial support from the grant DE-SC0021075 funded by the U.S. Department of Energy, Office of Science. Part of this work was performed at the Stanford Nano Shared Facilities (SNSF), supported by the National Science Foundation under award ECCS-1542152. M.T.K. acknowledges the National Defense and Science Engineering Graduate Fellowship. Z.H.A. and Y.W.Z. gratefully acknowledge the use of computing resources at the A*STAR Computational Resource Centre and National Supercomputer Centre, Singapore and the financial support from the Singapore A*STAR under grant AMDM A1898b0043. Y.W.Z. acknowledges the support from Singapore A*STAR SERC CRF Award.

**Supplementary Information for**

Extraordinary Strain Hardening from Dislocation Loops in Defect-Free Al Nanocubes

Mehrdad T Kiani, Zachary H Aitken, Abhinav Parakh, Yong-Wei Zhang, X Wendy Gu

Corresponding Author: X Wendy Gu

Email: xwgu@stanford.edu

**This PDF file includes:**

Figures S1 to S2

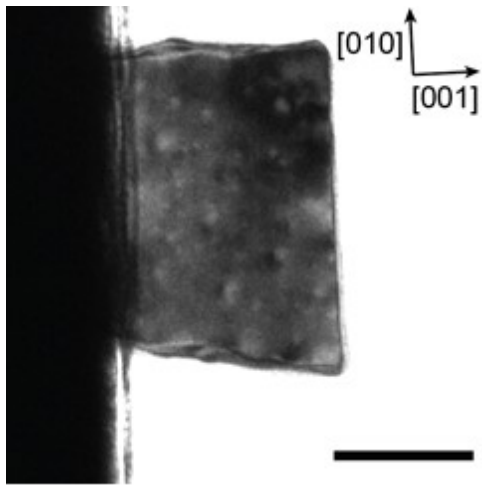

**Fig. S1.** Compressed Al nanocube imaged in the <100> direction. Scale bar is 50 nm.

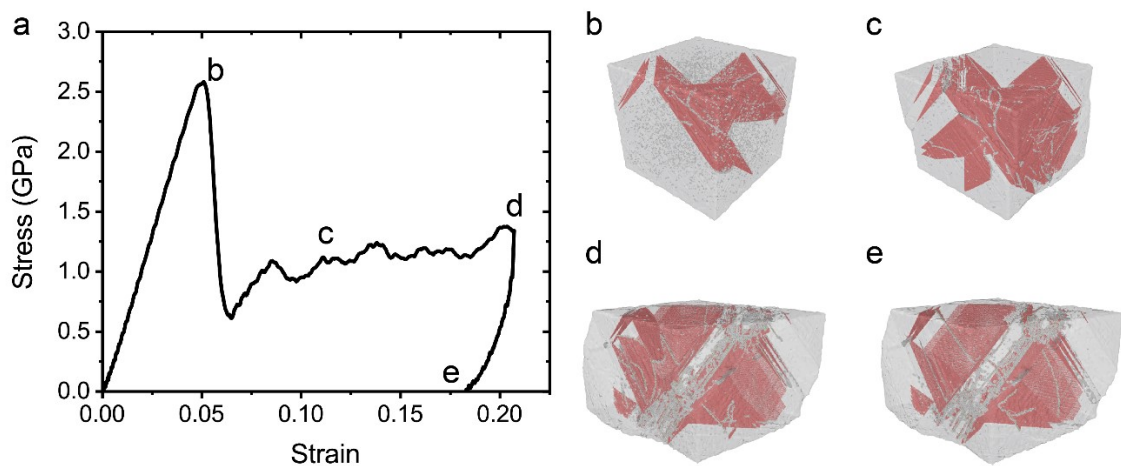

**Fig. S2.** Simulation results from compression of smooth nanocube without a rigid surface. a) Stress-strain data. Atomistic snapshots of deformed structure at strains of b) 0.05, c) 0.11, d) 0.21, and e) after unloading. Only non-FCC atoms are shown (i.e. defected or surface atoms), with specific CNA classification of red (HCP), blue (BCC), and white (unknown).